\documentclass{osa-article}

\journal{ome}

\articletype{Research Article}
\usepackage{float}

\begin{document}

\title{3D Imaging of Gems and Minerals by Multiphoton Microscopy}

\author{Benjamin Cromey\authormark{1,*}, Ryan J. Knox\authormark{1}, and Khanh Kieu\authormark{1}}

\address{\authormark{1}College of Optical Sciences, the University of Arizona, 1630 E University Blvd, Tucson, AZ 85716, USA}

\email{\authormark{*}bcromey@email.arizona.edu} 



\begin{abstract*}
Many optical approaches have been used to examine the composition and structure of gemstones, both recently and throughout history. The nonlinear optical behavior of different gemstones has not been investigated, and the higher order terms to the refractive index represent an unused tool for qualifying and examining a stone. We have used a multiphoton microscope to examine the nonlinear optical properties of 36 different gemstones and demonstrate that it is a useful tool for imaging them three-dimensionally up to the millimeter scale below the sample surface. The polarization dependence of second harmonic generation signals was used to examine the crystal orientations inside the minerals.
\end{abstract*}

\bibliography{sample}

\section{Introduction}
The science of examining and qualifying gemstones has used optics for centuries. In a lecture delivered at the Imperial Institute of Great Britain in 1895, Sir Henry Meirs, a mineralogist, discussed the clear advantages of the use of optics for the analysis of gemstones compared to existing traditional methods\cite{miers_precious_1895}. Non-contact and nondestructive measuring methods are vital with such valuable samples. However, the common tests of the era included scratching samples to check their hardness, weighing the samples in different media to find their specific gravity, and chemical analyses, which necessitated destroying part of a sample. By contrast, Sir Meirs argued for using conventional and polarized light microscopy, and measuring refractive index and absorption as more accurate, less destructive methods. Modern versions of many of his proposed methods are still commonly used by geologists today as ready tools for qualifying and identifying stones\cite{pirard_particle_2007,turner_reflectance_2017}.

In recent years, a number of more advanced imaging and analysis techniques have been used to examine gemstones. Raman spectroscopy can  identify minerals via their characteristic vibrational modes, which has been widely reported and used in the literature\cite{bersani_applications_2010,hope_raman_2001,reiche_situ_????,barone_red_????,kiefert_identification_2000,giarola_raman_????,culka_gem_2016}. This approach lacks the ability to resolve structural details, however, since each Raman spectrum must be taken in a point-by-point manner. Terahertz spectroscopy has been used to analyze gems, but without any high resolution or structural information\cite{yu_identification_2010,han_lattice_2015}. Electron microscopy can resolve high resolution structural details down to the atomic scale, but is not practical for examining a large scale portion of a sample, and cannot probe deep beneath a sample surface\cite{krivanek_atom-by-atom_2010}. X-ray micro-CT can resolve structural details in 3D, but has a long acquisition time and requires expensive equipment\cite{sahoo_surface_2016}. Proton Induced X-ray Emission (PIXE) can perform an elemental analysis of a gemstone, but cannot provide any structural information\cite{venkateswara_rao_trace_2013}. The Atomic Force Microscope can resolve incredibly fine details on the surface of a gemstone, and even examine the pyroelectric nature of certain stones, but cannot obtain any information from below the surface\cite{afm}.

The first introduction of the multiphoton microscope focused primarily on its usefulness in the biological sciences\cite{denk_two-photon_1990}. Recently, the multiphoton microscope has been shown to be very useful in the characterization of materials as well. We have designed and used several multiphoton microscopes for a variety of applications\cite{Kieu:13,armpm}. We have used them to examine the layers and grain boundaries of Molybdenum Di-sulphide\cite{mos2_2,mos2}, rapidly characterize large sections of graphene\cite{graphene}, and utilize second and third harmonic generation to investigate the layers of Galium Selenide\cite{GalSel}. In this publication, we demonstrate the utility of the multiphoton microscope (MPM) in the study of gems and minerals. The MPM can capture the beautiful structural details of these stones non-destructively with three dimensional sub-micron resolution through depths on the millimeter scale. We hope that this will be a new useful tool that will enable mineralogists to determine structural details beneath the surface, as well as crystalline axis orientation by adjusting the laser polarization. This new information obtained using the MPM may provide useful insights into the formation of these gemstones. Lastly, gems and minerals have long been valued due to their beauty and rarity. We feel that this tool is a new way to appreciate them, by revealing their hidden details beneath the surface. 

\section{Methodology}

\subsection{Multiphoton Microscope Description}

The multiphoton microscope used was designed and built in-house, and is controlled by LabVIEW software created by our research group. Since its design is described elsewhere in the literature\cite{Kieu:13}, only a brief description will be given here. The multiphoton uses nonlinear optical interactions to create contrast in a sample. A pulsed laser source with very high peak power, but relatively low average power, is raster scanned by a pair of galvanometric scan mirrors, and focused onto the sample with a microscope objective, creating an image in a point-by-point fashion. The high peak power (kilowatt level) of the laser pulses is ideal for creating strong nonlinear signals from the samples. However, the low average power (milliwatt level) prevents the samples from being damaged. Our system uses a dichroic mirror to split the signal to two photomultiplier tubes (PMTS), enabling the simultaneous detection of two different signals. In this project, the microscope was configured to image with Second Harmonic Generation (SHG) and Third harmonic Generation (THG).

Two different lasers were used to image the samples, both of which were femtosecond pulsed fiber lasers, one at a wavelength of 1040nm  and the other at 1560nm. Both lasers have an 8 MHz rep rate, pulse widths on the 100 femtosecond scale, and average powers around 70 mW. The samples were imaged with both lasers in order to examine the different information and different depths that could be probed by each source. To split the signal light from the 1040nm and 1560nm lasers into the two detection channels, 414nm and 560nm dichroic mirrors were used, respectively. For capturing images, a Nikon 20x .75NA objective was used.  This objective was chosen as the best balance of resolution, field of view, and working distance. The resolution for this objective was calculated according to equations published in the literature\cite{Zipfel2003}. For the 1040 laser, the diffraction limited lateral resolution was 520 nm for SHG, and 425 nm for THG. For the 1550 laser, the lateral resolution was 770 nm for SHG, and 630 nm for THG.  Since the small working distance associated with higher NA objectives was not an issue for most images of these samples, the Nikon 20x .75NA objective was used for every image captured unless otherwise noted.  For the deeper images where a longer working distance was needed, the objective was changed to a 20x .5 NA aspheric lens, with the trade off in a decrease in resolution (about a factor of 1.5 times lower resolution). The THG and SHG signals were collected by the same microscope objective in an epi-detection imaging scheme. 

\subsection{Sample Description and Imaging Methodology}

\begin{figure}[H]
\begin{center}
\includegraphics[width=4.5in]{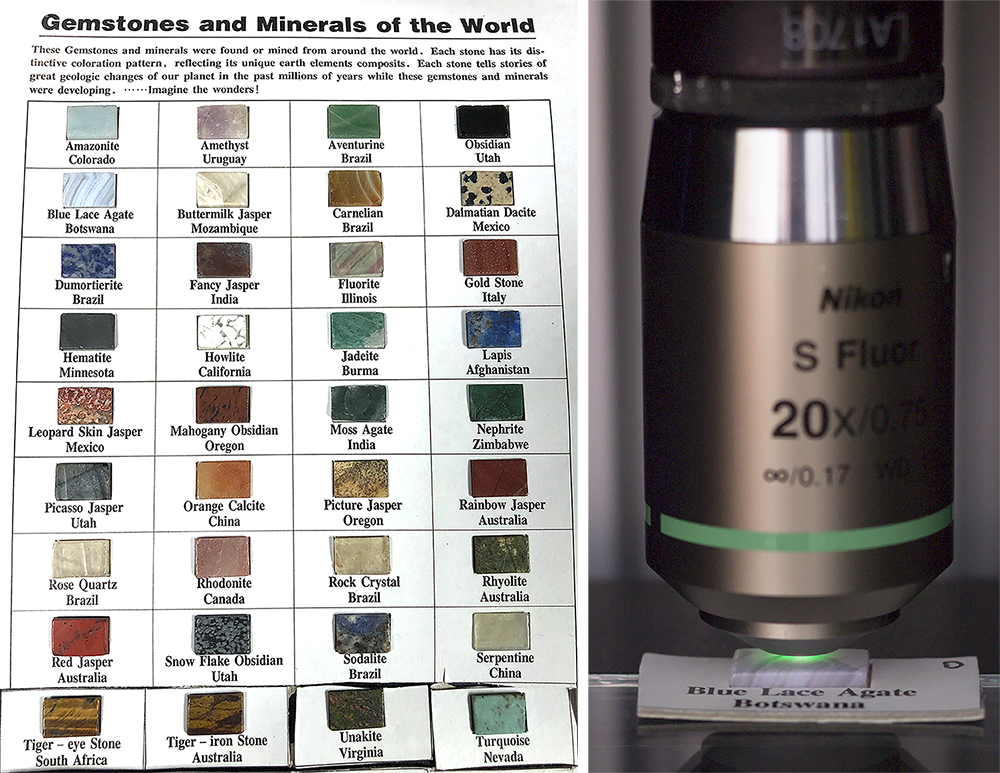}
\caption{Left: Gemstone Sample Collection. The individual gemstones were cut out of the large cardboard backing to fit underneath the microscope objective. This can be seen in the bottom row of the figure. The names and locations of each mineral were labeled by the vendor. Right: The Blue Lace Agate sample underneath the microscope. This sample had particularly strong SHG, as can be seen from the green glow from the sample (excitation laser was at 1040 nm).}
\label{gems_collection}
\end{center}
\end{figure}

A collection of 36 gem stones (10mm tall by 15 mm wide, 2mm thick), seen in Fig. \ref{gems_collection} on the left, were purchased at the Tucson Gem and Mineral Show, a world famous event to the gem community\cite{GemandMineralShow}.

Each gem was cut out of the large cardboard piece so the samples could easily fit under the objective, shown on the right of Fig. \ref{gems_collection}. Each gem was first imaged using the 1040 nm laser, and regions of interest were examined with SHG and THG. Some of these signals were strong enough to be seen by the naked eye, as was the case of the Blue Lace Agate sample. The second harmonic of 1040nm is 520nm, a green wavelength, which can be clearly seen on the right of Fig. \ref{gems_collection}. These signals were isolated using bandpass filters, specifically the 340/22 filter for THG and a 517/20 filter for SHG (Semrock). The two signals split to the two detection channels by a 414 nm dichroic (Semrock). The depth limit to which a sample could be imaged through focus was also examined for each sample. This was done with a set of through focus images, commonly called a Zstack.  Each Zstack was set to take an image every few microns until the signal became too weak to produce a useful image.  The 1040nm laser was then swapped for the 1560nm laser, and the 414nm dichroic swapped for a 560nm dichroic. The THG was isolated by a 517/20 filter, and the SHG by a 780/12 filter (Semrock).  

\section{Experiment Results}
\label{Results}

\begin{figure}[H]
\begin{center}
\includegraphics[width=4.25in]{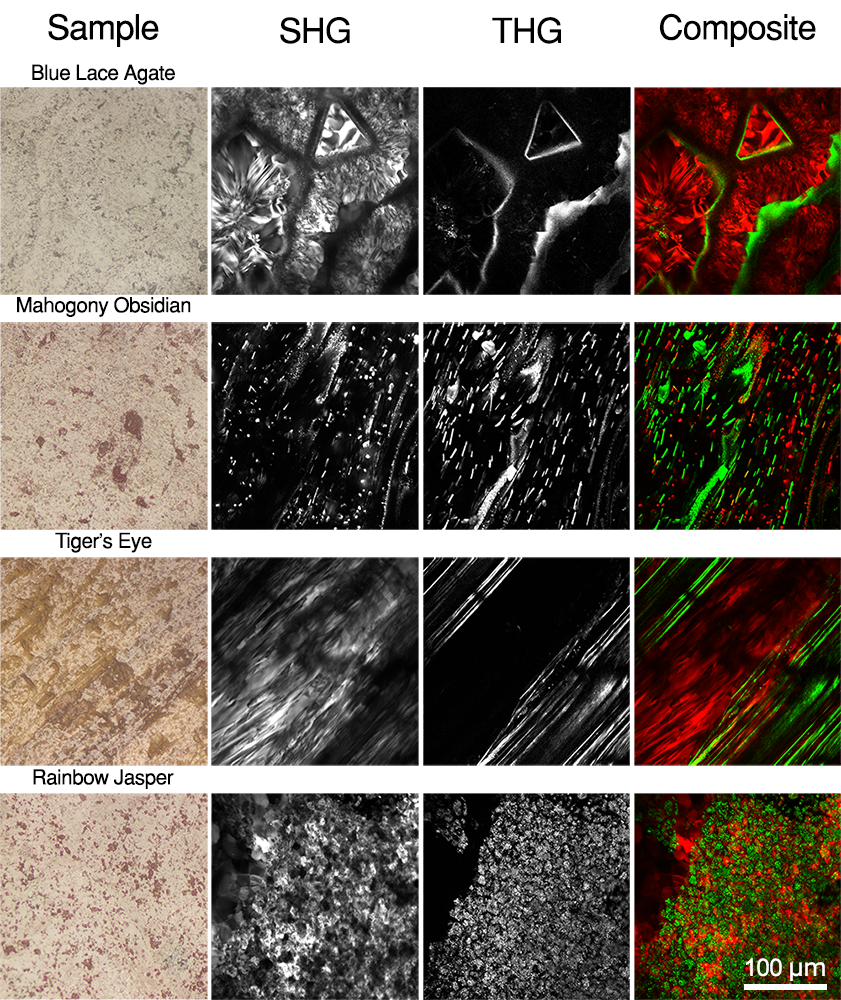}
\caption{A collection of image results from some of the more interesting looking samples. Left: An optical microscope image. Center left: SHG information. Center Right: THG information. Right: Composite image with SHG red and THG green. Samples were imaged with the 1040 nm laser, and MPM images were taken $\approx$100 $\mu m$ below the surface (See Visualization 1 for through focus video of the first sample, Blue Lace Agate). Each MPM image has the same scale of 250 $\mu m$ across, with a scale bar shown on the bottom right.}
\label{results}
\end{center}
\end{figure}

A selected set of images from varying depths are displayed in Fig. \ref{results}. The images were captured by our lab created software, and processed using the Fiji release of ImageJ\cite{Fiji}.  In the four column figure, the image on the left is from a standard white light microscope (Zeiss Axioplan 2), followed by an image that shows regions of SHG, then regions of THG. The last column combines the two signals together with false color, with SHG colored red and THG colored green. Only small amounts of image adjustments were made, including adjusting the brightness of the two channels. All multiphoton images shown in Fig. \ref{results} were captured using the 1040 nm laser and the 20x .75 NA objective.

In these images, all of which were taken below the surface of the minerals, we can begin to see how THG and SHG highlights very different features in these stones. While the optical microscope images look relatively similar, the nonlinear signals from each stone vary widely in both SHG and THG. SHG is generated from a crystal structure with broken symmetry, or from structures without centro-symmetry. THG is generated from boundaries between regions of differing Third Order Nonlinear response, or $\chi_3$. THG is not generated in a bulk uniform medium due to the Gouy phase shift through the focal spot\cite{Boyd,SHGandTHGtheory}. In the Blue Lace Agate, SHG shows fine structural variations within the stone, while THG highlights the boundaries between crystalline regions (See Visualization 1 for a through-focus video). In the Mahogany Obsidian, very different structures are highlighted between SHG and THG. In the Tiger's eye, some bulk SHG is seen, while the THG highlights the fine line structure that gives Tiger's eye its characteristic appearance. In the Rainbow Jasper, both SHG and THG contain fine structural information, enough that the composite image is difficult to read. The information gathered by the microscope represents sub-micron resolved three dimensional information that could be a crucial analysis tool to a mineralogist looking to examine the components of a composite stone.  The signal from many of these samples were very strong. Compared to imaging a piece of pure Quartz, which has very well-known nonlinear coefficients \cite{quartz1,quartz2}, many of the SHG signals from the gemstones were measured to be hundreds of times stronger. Since there is insufficient room to report images from all 36 samples in this paper, images from every sample can be found at \url{wp.optics.arizona.edu/kkieu/gemstone-images/}.

An Ocean Optics QE65000 spectrometer was used to confirm the signals coming off of these samples when imaged with the 1560 laser. The sample spectra can be seen in Fig. \ref{Spectra} below. THG appears around 517 nm, and SHG around 780 nm. Because of the wide spectrum of the mode locked laser, shown in the inset of Fig. \ref{Spectra}, there are multiple peaks to the SHG and THG spectra. 

\begin{figure}[H]
\begin{center}
\includegraphics[width=4.5in]{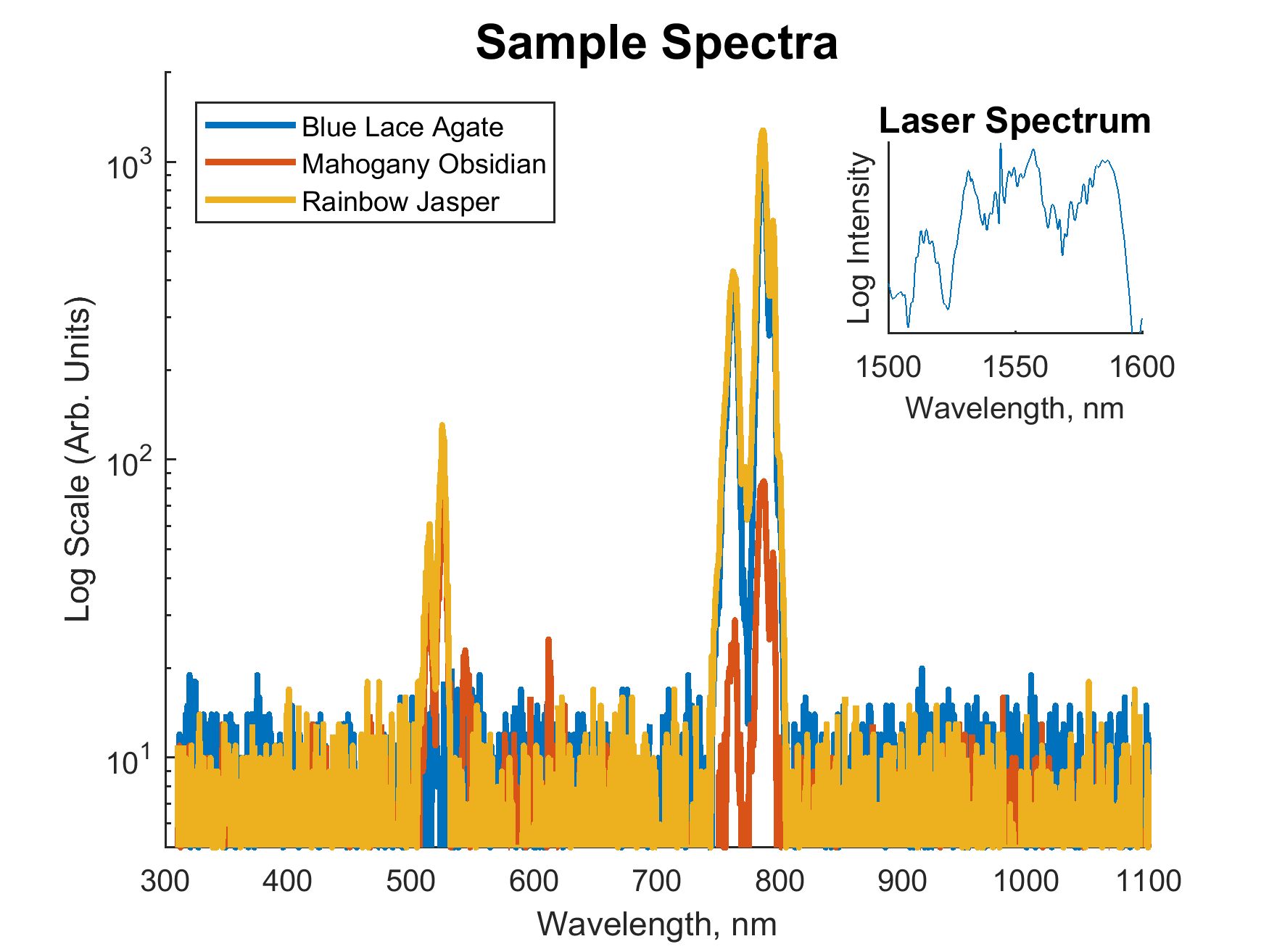}
\caption{Signal spectra from selected samples are displayed. Most samples predominantly showed strong SHG and relatively strong THG. For comparison, the inset shows the spectra from the laser to create this data.}
\label{Spectra}
\end{center}
\end{figure}

Interestingly, we do not see any fluorescence from these stones, even from the fluorite sample in our collection. It is possible that the excitation is too long at 1560 nm and 1040 nm to get efficient excitation of any fluorphores which may be present within the stone. It is possible that a shorter wavelength femtosecond source such as a Ti:Sa laser would be able to excite them.

\subsection{3D Imaging Results}

\begin{figure}[H]
\begin{center}
\includegraphics[width=4.25in]{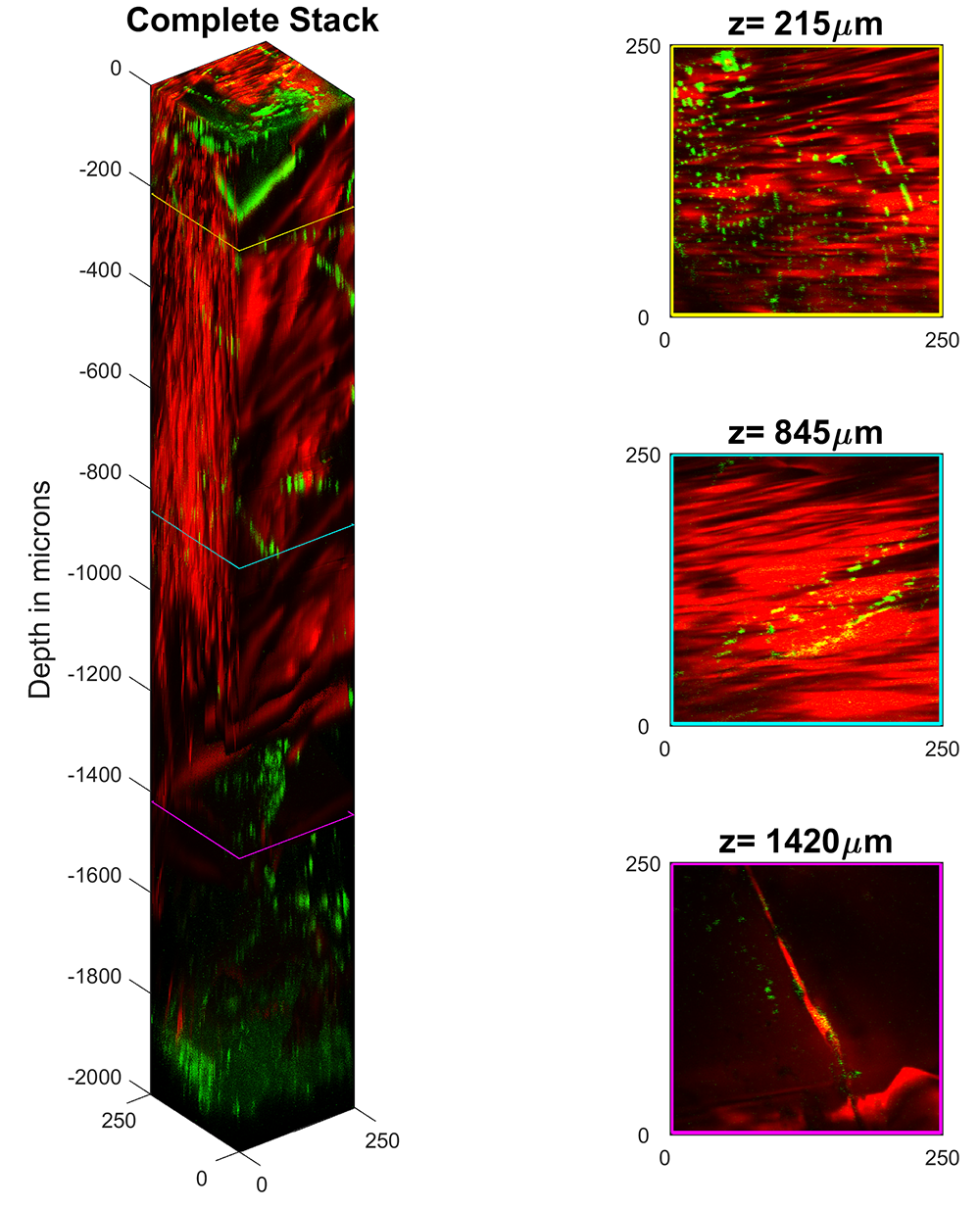}
\caption{A Zstack of images from the Amethyst sample, taken with a 20x .5 NA objective and the 1560 nm laser. On the left, all of the images taken in the stack are displayed as a function of depth. On the right, three slices are displayed separately, highlighting different depths. The depths can be seen in the figure on the left as highlighted in Yellow, Cyan, and Purple respectively. As in the other figures, green shows information from THG, and red shows information from SHG. All numbered distances are in microns. See Visualization 2 for a through focus video.}
\label{Amethyst_zstack}
\end{center}
\end{figure}

The sample that could be imaged the deepest was the Amethyst sample at the 1560 nm wavelength. An objective with a longer working distance was required, so the 20x .5 NA objective was used. The sample was 2 mm thick, and the microscope was able to image all the way through it, as seen in Fig. \ref{Amethyst_zstack}. Several highlighted depths are shown on the right side of the figure. 

This sample appeared to be the most clear to the eye, so it was not surprising that it allowed the deepest penetration. Because of this clearness, it is difficult to discern information at depth with visual techniques like light microscopy, since there is insufficient contrast. However, the MPM can see interesting structural information throughout the sample, predominantly in SHG, but smaller dot regions, possibly impurities, can be detected clearly with THG through the entire depth as well.

\subsection{Polarization Dependence of Signal}
\label{polar}

\begin{figure}[H]
\begin{center}
\includegraphics[width=4.5in]{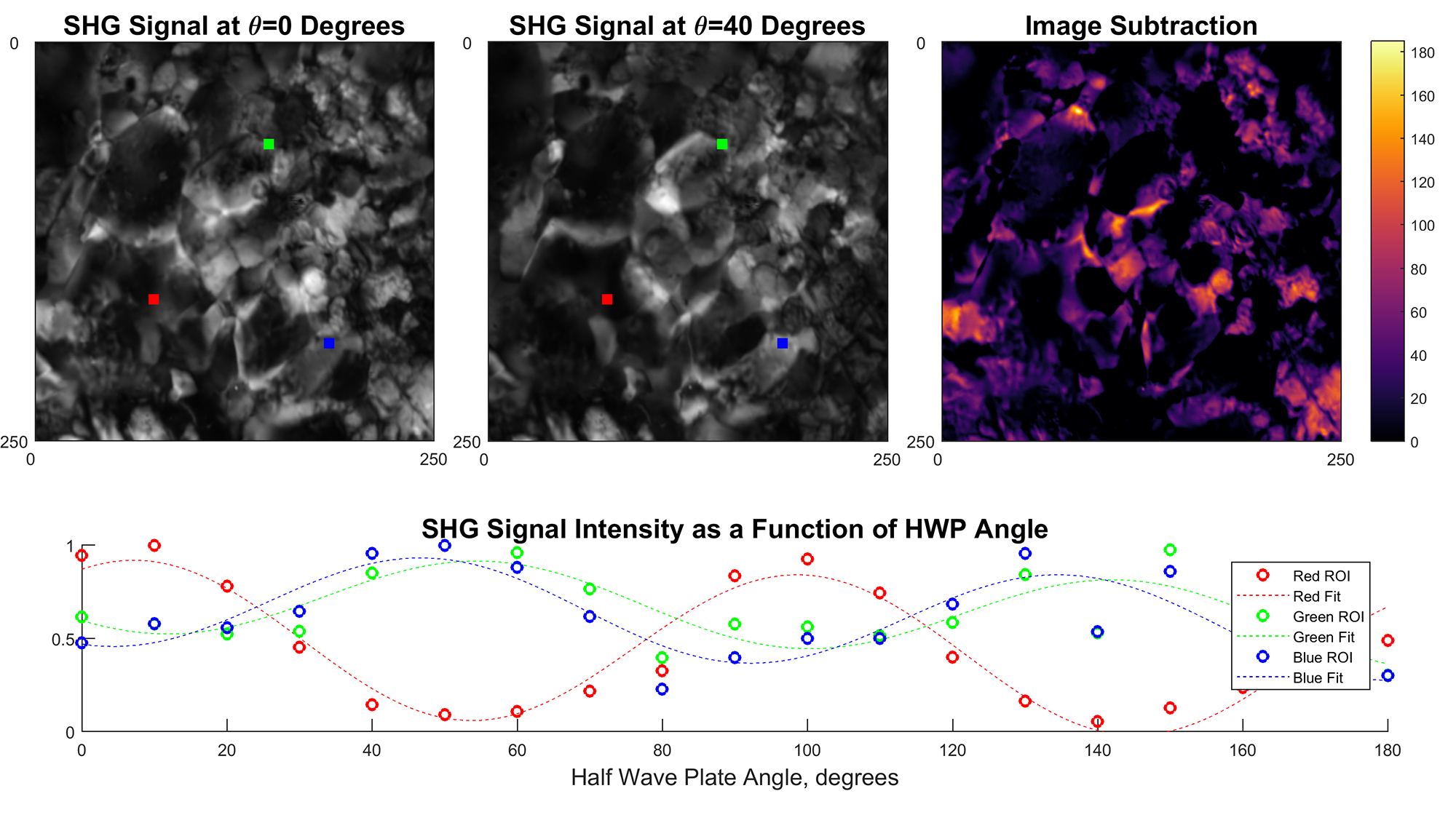}
\caption{Top left: SHG signal with a HWP angle of $0^\circ$. Top Center: SHG signal with a HWP angle of $40^\circ$. Top Right: The first two images are subtracted, and the difference is displayed, highlighting regions where the signal changed significantly. A scale bar is displayed for the intensity of each pixel. The dimensions on each picture are in microns. Bottom: the signal intensity at three regions of interest (ROI) are plotted as a function of angle. See Visualization 3 for a video showing the change in signal as a function of polarization angle.}
\label{shgangle}
\end{center}
\end{figure}

The second order nonlinear response $\chi_2$ is generally polarization dependent\cite{Boyd}. The fiber used in the 1040 nm laser is not polarization maintaining (PM), so the polarization of the femtosecond pulses was elliptical. However, the 1560 laser uses PM fiber with a linearly polarized output, and was used to examine the different information gathered from adjusting the polarization angle hitting the sample. To test SHG signal intensity at different polarization angles, a half wave plate (HWP) was rotated from 0 to 180 degrees while capturing an image every 10 degrees.  The intensity vs HWP angle was extracted from the images and plotted, shown in Fig. \ref{shgangle}, alongside several images. The HWP angles are labeled in each portion of the figure.  A short movie was then created showing the changes in intensity for the Dumortierite sample as the polarization angle changed (Visualization 3). 

Two different images at different polarization angles, top left and top center in Fig. \ref{shgangle}, were subtracted, creating the image in the top right. This highlights regions where the SHG has changed between the two polarization states. The signal was plotted for every picture taken at three highlighted regions, seen in Red, Green, and Blue in the figures and plotted in those colors, respectively. Each plot was fit to a sinusoidal function. It can be seen that some portions of the sample have signal that moves in roughly the same way as a function of angle, while other regions move separately. These are regions of differing crystal axes. This angularly dependent SHG signal strength information can be used to determine the local crystalline axes\cite{crystalSHG}. We can see that the red region has its maximum signal around a HWP angle of ~10 degrees, with another peak 90 degrees later, which makes sense because of the angle doubling nature of a HWP. We can see similar information for the Blue and Green points, which share approximately the same crystalline axis direction. 

\section{Results and Discussion}

The current acquisition time for these images on our microscopes can be relatively slow. For example, the Zstack shown in Fig. \ref{Amethyst_zstack} took a little over four hours to capture, since one picture was captured every micron increment in z, leading to two thousand images that were created for both SHG and THG. However, this is not a limit of this technological approach. Our multiphoton microscope is optimized for compactness and cost, not speed. Resonant galvano mirrors or acoustic-optic deflectors can decrease the rastering time significantly, allowing a much shorter acquisition time\cite{otsu_optical_2008}. Alternatively, a high sensitivity CCD can be used to remove the need to raster altogether, allowing much higher frame rates\cite{Bewersdorf:98}.

The depth to which samples could be imaged varied from a few hundred microns to the 2 mm depth reported in Fig. \ref{Amethyst_zstack}. Not surprisingly, the depth also depended on the wavelength used. Typically, the longer 1560nm wavelength allowed for approximately twice as much imaging depth, which makes sense if Rayleigh scattering is the dominant process. The different minerals certainly have different absorption characteristics, which played a role in limiting the imaging depth as well. Typically, the samples allowed for imaging between 100-300 microns of depth for the 1040 nm laser, and 200-600 microns for the 1550 laser. Future work could focus on measuring the absorption and scattering coefficients of these samples from the zstacks that were captured, as well as determining the nonlinear coefficients of the refractive index. 

Many of the samples exhibited interesting variations in signal as a function of polarization angle for the 1550 laser. Our approach will allow for the determination of the crystalline axis at any depth. The images and approach demonstrated in Section \ref{polar} do this for a 2D projection of the crystalline axis, but this could be extended to a 3D measurement with multiple depths and multiple images at different polarizations. 

As discussed, many of the samples had very strong nonlinearities. The strong signals observed came from naturally created structures grown with no thought to phase matching or any other efficiency concern. Some of these natural stones could be used for harmonic generation very inexpensively

\section{Conclusion}

This work has demonstrated the utility of the three dimensional imaging capability of the multiphoton microscope. Throughout the 36 samples examined, the multiphoton was able to discern crystalline boundaries, sub-micron features, and other structural information. SHG enabled the determination of crystalline axes, a useful structural and analysis feature. We believe that the multiphoton microscope will yield valuable information to mineralogists and geologists, and with our instrument have shown that it can be done affordably and compactly. 

\section*{Funding}
National Science Foundation Graduate Research Fellowship under DGE-1143953, NSF ECCS under Grant Number \#1610048.

\section*{Disclosures}
The authors declare that there are no conflicts of interest related to this article.


\end{document}